\documentclass[conference]{IEEEtran}
\IEEEoverridecommandlockouts

\IEEEpubid{%
   \makebox[\columnwidth]{ISBN 978-3-903176-39-3~\copyright~2021 IFIP \hfill}%
   \hspace{\columnsep}%
   \makebox[\columnwidth]{ }%
}
\usepackage{cite}
\usepackage{graphicx}
\usepackage{listings}
\def\BibTeX{{\rm B\kern-.05em{\sc i\kern-.025em b}\kern-.08em
    T\kern-.1667em\lower.7ex\hbox{E}\kern-.125emX}}

\begin{document}

\title{Enabling self-verifiable mutable content items in IPFS using Decentralized Identifiers}

\author{\IEEEauthorblockN{Nikos Fotiou, Vasilios A. Siris, George C. Polyzos}
\IEEEauthorblockA{Mobile Multimedia Laboratory,\\
Department of Informatics School of Information Sciences and Technology\\
Athens University of Economics and Business, Greece\\
\{fotiou,vsiris,polyzos\}@aueb.gr}
}

\maketitle

\begin{abstract}
In IPFS content identifiers are constructed based on the item's data therefore the binding between an item's identifier
and its data can be deterministically verified. Nevertheless, once an item is modified, its identifier also
changes. Therefore when it comes to mutable content there is a need for keeping track of the 
``latest'' IPFS identifier. This is achieved using naming protocols on top of IPFS, such as IPNS and DNSlink,
that map a constant name to an IPFS identifier, allowing at the same time content owners to update these mappings.
Nevertheless, IPNS relies on a cryptographic key pair that cannot be rotated, and DNSlink does not provide content authenticity protection.
In this paper, we propose a naming protocol that combines DNSlink and decentralized identifiers to
enable self-verifiable content items. Our protocol provides content authenticity
without imposing any security requirement to DNSlink. Furthermore, our protocol prevent fake content even if
attackers have access to the DNS server of the content owner or have access to the content
owner secret keys. Our proof of concept implementation shows that our protocol is feasible and can be used
with existing IPFS tools. 
\end{abstract}

\begin{IEEEkeywords}
decentralization, delegation, privacy, self-sovereignty. 
\end{IEEEkeywords}

\section{Introduction}
Modern communication systems often try to provide content \textbf{integrity} and \textbf{authenticity} protection.
By protecting content integrity, a content item recipient can verify that the received item \emph{has not been modified
during transmission}, whereas by protecting authenticity, a recipient can verify that the received item \emph{is indeed the
requested item}. Although content integrity can be easily provided, content authenticity protection is a more 
challenging problem that even commonly used communication systems, sometimes, fail to provide (e.g., Apple's iMessage
lack of content authenticity verification allowed attackers to send messages on behalf of legitimate users~\cite{Gar2016}).

The InterPlanetary File System (IPFS)~\cite{Ben2014} achieves content integrity and authenticity protection by using
content identifiers (CIDs) based on the content's cryptographic hash. A user can request a content item
from the IPFS network by providing the item's CID; then she can easily verify that the received item is indeed 
the requested one simply by applying the same hashing algorithm. Nevertheless this approach has the drawback 
that whenever an item is modified, its CID changes also. In order to support mutable items IPFS natively supports two 
``overlay'' naming solutions that allow users to find the ``current'' CID of an item: the   
InterPlanetary Name System (IPNS)~\cite{ipns} and DNSlink~\cite{dnslink}.

\subsection{InterPlanetary Name System (IPNS)}
The InterPlanetary Name System (IPNS) uses public key digests as content names. An IPNS name is mapped to a record
that includes ``information'' that can be used for retrieving an item from the
IPFS network. Such information could be the CID of the item, another IPNS name, or a DNSlink name
(we discuss DNSlink names in the following subsection): this information is referred to as the item's \emph{address}. 
These records are stored inside the IPFS network,
they can be retrieved using standard IPFS tools, and they can be modified by the holder of the
corresponding private key (e.g., in cases they concern mutable items). 

IPNS provides content authenticity by signing its records with the private key that corresponds to the IPNS name. On the
other hand  IPNS does not allow key rotation; content owners should
maintain a key pair for each IPNS name: the hash of the public key is the name and the private key is used for
signing the associated record. In case the private key is lost or breached, the 
content owner loses control of the corresponding record. Furthermore,
this limitation of IPNS impedes content storage delegation. Consider for example the case of a
service that `hosts' dynamic web pages in IPFS using IPNS. This service would be responsible
for keeping up-to-date the corresponding IPNS record. If the site owner decides to change hosting
service she cannot be sure that a) the current hosting service will reveal to the site owner the 
private key associated with the IPNS name, and b) the current hosting service will erase that
private key.   

\subsection{DNSlink}
DNSlink uses domain names, usually prefixed with the string ``\_dnslink'' as content names. A DNS record
for a DNSlink name is an 
ordinary TXT record that includes an item's \emph{address}.
Even if the address is an IPFS CID, DNSlink can be still 
used for mutable items, providing that every time an item is modified the 
DNS record of its DNSlink name is updated accordingly. Therefore, the retrieval of an item based on a DNSlink name
requires at least a DNS resolution. 

\begin{figure}
  \centering
  \includegraphics[width=0.9\linewidth]{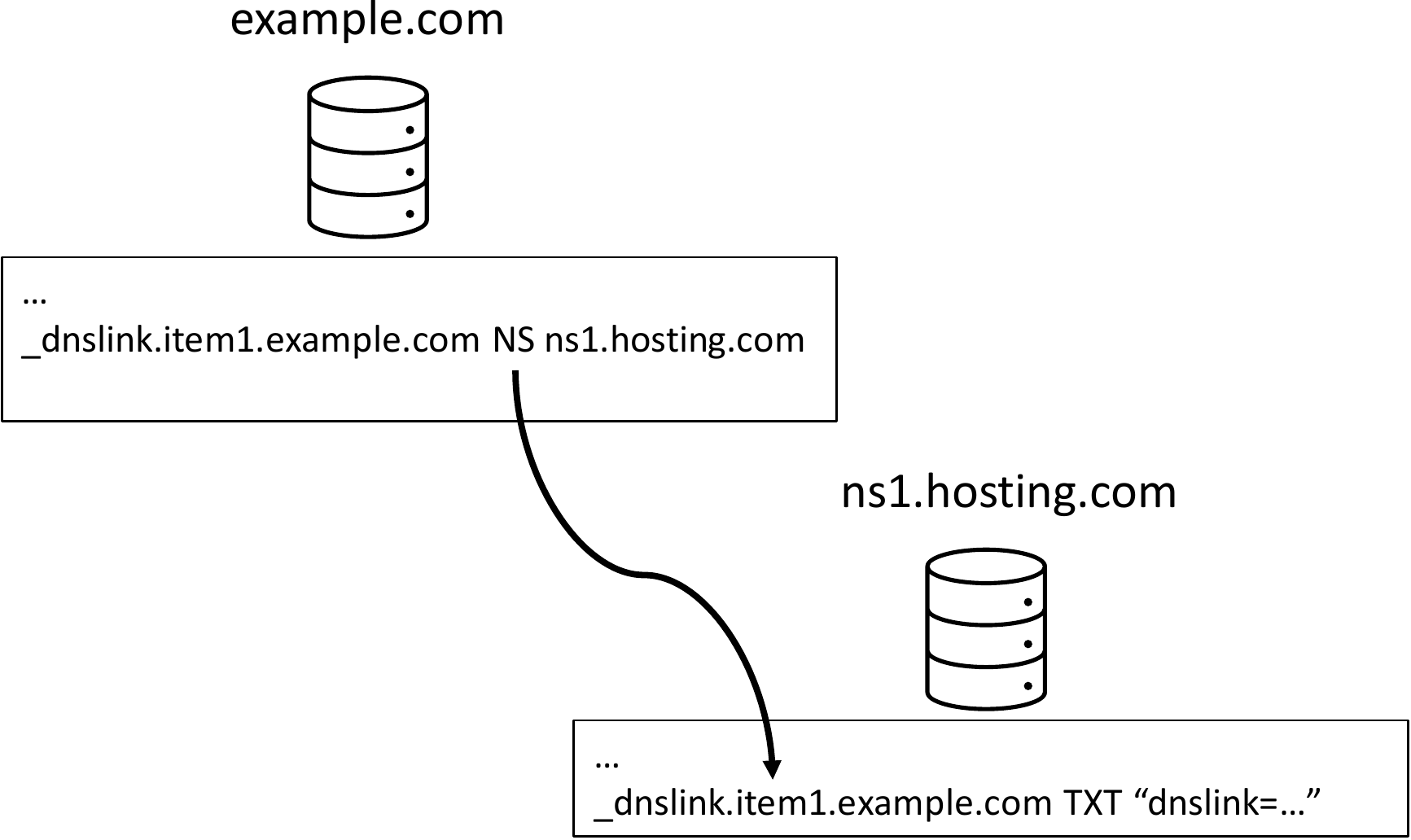}
  \caption{Delegation using DNSlink.}
  \label{fig:del}
\end{figure}

DNSlink can take advantage of all features of DNS in order to improve content availability (e.g., cached DNS records,
multiple DNS server, use of 3rd party DNS services with greater availability, and others). 
Moreover, content delegation can be easily achieved by adding a Name Server (NS) record for the
DNSlink (sub)domain in the content owner's DNS server, which will point to a DNS server owned by the 
delegatee: in the hosting service scenario, the hosting service DNS will become the Name Server of the
corresponding DNS record; when the site owner decides to change hosting service she can
simply modify the DNSlink domain record in her DNS server to point the DNS server owned by 
the new hosting service. An example of this process is illustrated in Figure~\ref{fig:del}.
In this example, content owner ``example.com'' has delegated the item ``item1.example.com'' to the
hosting service ``hosting.com''. For this reason the corresponding record in the ``example.com''
DNS server maps to a Name Server that belongs to ``hosting.com''.

On the other hand, DNSlink relies on the security of DNS to provide content authenticity: an attacker
can modify the mapping from a DNSlink name to the actual items stored in the
IPFS network by either having access to the DNS server where the DNSlink record is stored, or 
by intervening to the DNS resolution process. 

\subsection{Contributions}
In this paper we propose a solution that uses \emph{decentralized identifiers} (DIDs) as content names and leverages
DNSlink to map a DID to an IPFS CID. A DID is new type of self-administered, globally unique identifier~\cite{did-primer}. 
Due to the intriguing security and privacy characteristics of the DID paradigm,
many research efforts investigate the potential of using DIDs for improving the security and privacy of emerging
technologies. Chadwick et al.~\cite{Cha2019} propose the integration of DIDs and
Verifiable Credentials with FIDO Universal Authentication Framework (UAF) in order
to provide safer and more private online account management. 
DIDs are also considered for improving the security and privacy of IoT
systems (e.g.,~\cite{Ans2019},\cite{Fed2020},\cite{Kor2019},\cite{Ter2020}). Davie et al.~\cite{Dav2019} propose
a four-layer architectural stack, based on DIDs, for establishing trust between peers over 
the Internet and other digital networks. Lagutin et al.~\cite{Lag2019} investigate
the application of DIDs and Verifiable Credentials into the OAuth 2.0 authorization process. 
Finally, Munoz~\cite{Mun2019} discusses the
advantages of an integration of eIDAS and DIDs. 

A DID can be treated as an opaque string which is
associated with a public-private key pair.
The latter key pair is used in our solution for signing (and verifying) some content item \emph{metadata},
which are then used as a proof of authenticity. 
Compared to DNSlink, our solution achieves content authenticity protection 
without making any assumption about the security of DNS, by sacrificing however human readability.
As opposed to IPNS, signed records in our solution are attached to the items
themselves, hence items can be retrieved as long as the DNS service is available. Additionally, and contrary
to IPNS, the signing key pair can be rotated, or even re-used for many items, without needing
to change the corresponding DID (which is used as the name of the mutable item).
The latter property facilitates key management and makes our solution more resilient
to key breaches. Finally, our solution enables content storage delegations and it
can be used with existing IPFS tools. 

The remainder of this paper is organized as follows. In Section 2 we introduce DIDs and we
detail our DID-based naming system. In Section 3 we present implementation details of our
solution and we analyze its security properties. Finally, in Section 4 we discuss future
extensions of our work and we conclude our paper. 

\section{System Design}
\subsection{Decentralized Identifiers and the did:self method}
Decentralized Identifiers (DIDs), defined by W3C, are a new type of globally unique identifier 
designed to enable individuals and organizations to generate their own identifiers using systems they trust~\cite{did-spec}.

A DID architecture can be regarded as a key-value lookup system, where the key is 
the Decentralized Identifier (DID) and the value is a DID \emph{document}.
A DID document includes, among other things, public keys that can be used
as \emph{verification methods}. For example, a DID document may include
public keys that can be used for authenticating the DID \emph{owner}, public keys that can be used
for verifying digital signatures generated by the owner, and other related information. 
Usually, a DID document is maintained by a DID registry which is responsible for
implementing proper security and access control mechanisms. Registries allow
$3^{rd}$ parties to lookup DID documents and they provide \emph{proofs} of correctness (e.g., 
a proof can be a digital signature generated by the registry). 

DID specifications do not dictate the actual contents of a DID document, neither
define how a registry operates. Instead these are
left as design choices to individual DID instantiations, also referred to as DID \emph{methods}.
Our system uses a DID method known as \emph{did:self}~\cite{didself}. 

A key property of did:self is that it does not require any trusted registry. DID owners are responsible
for disseminating their DID documents by themselves, e.g., by directly transmitting them to interested
parties, or by storing them in a Web server: did:self assures that a DID document is `correct'
even if is retrieved over an unsecured channel. 

A did:self-based DID is a base64url~\cite{rfc4648} encoded Ed22519 public key~\cite{Ber2012} 
prefixed with the string ``did:self:'', e.g.,

\[did{:}self{:}6D0RjXZfW58v4DGt...Kzn9fghX94LvrMDxo\]

A \emph{DID document} in did:self is a JSON-encoded file that may include any of the properties defined
by the DID specifications. Our solution uses the following properties:

\begin{itemize}
    \item \texttt{id}: The DID which the document concerns.
    \item \texttt{assertion}: An  Ed22519 public key expressed using 
    the ``JsonWebKey2020'' notation~\cite{did-reg}. 
\end{itemize}

The \emph{assertion} key is used for generating digital signatures on behalf of the DID owner.
An example of a DID document is included in the following listing. As it can be seen, line 1
includes the DID, and lines 3-10 define the assertion key, which is in essence the JSON web key
representation~\cite{rfc7517} of an Ed22519 public key (lines 7-9). 

\begin{lstlisting} [caption={An example of a DID document used in our solution.},label={list:did}]
    {
      ``id": ``did:self:6varD0Rj...",
      ``assertion": [{
        ``id": ``#key1", 
        ``type": ``JsonWebKey2020", 
        ``publicKeyJwk": {
          ``crv": ``Ed25519", 
          ``x": ``7wJkufDc...", 
          ``kty": ``OKP"
        }
      }]
    }
\end{lstlisting}
Additionally, each DID document is associated with a \emph{proof}
which is a the compact serialization of a JSON Web Signature (JWS)~\cite{rfc7515}.
The payload of the proof is a JSON string that includes the following 
properties: 
\begin{itemize}
\item \texttt{id}: The DID.
\item \texttt{created}: The date and time when the proof was generated.
\item \texttt{expires}: An optional expiration time.
\item \texttt{sha-256}: The base64url encoded hash of the DID document, calculated using SHA-256.
\end{itemize}

The signature of the proof is generated using EdDSA and the private key that corresponds to the
DID. This proof is used for validating the binding between a DID document and the corresponding
did:self DID. In particular any entity can trivially verify that a DID document is valid by executing
the following steps:
\begin{enumerate}
    \item Verify that the DID is included in the \texttt{id} property of the proof.
    \item Verify that the digest of the DID document is the same as the \texttt{sha-256} property of the proof.
    \item If the \texttt{expires} property is set, make sure that the proof has not expired. 
    \item Verify the signature of the proof using did:self DID (it is reminded
    that a did:self DID is a public key).
\end{enumerate}

\subsection{Self-verifiable content items}
Using did:self we can construct self-verifiable content items in the sense that given an item name
and its data, any entity can verify whether or not these data are the ``real'' data of the item. This
verification is achieved without relying on any trusted third party. Such a self-verifiable content
item is constructed as follows.

\begin{figure}
  \centering
  \includegraphics[width=0.95\linewidth]{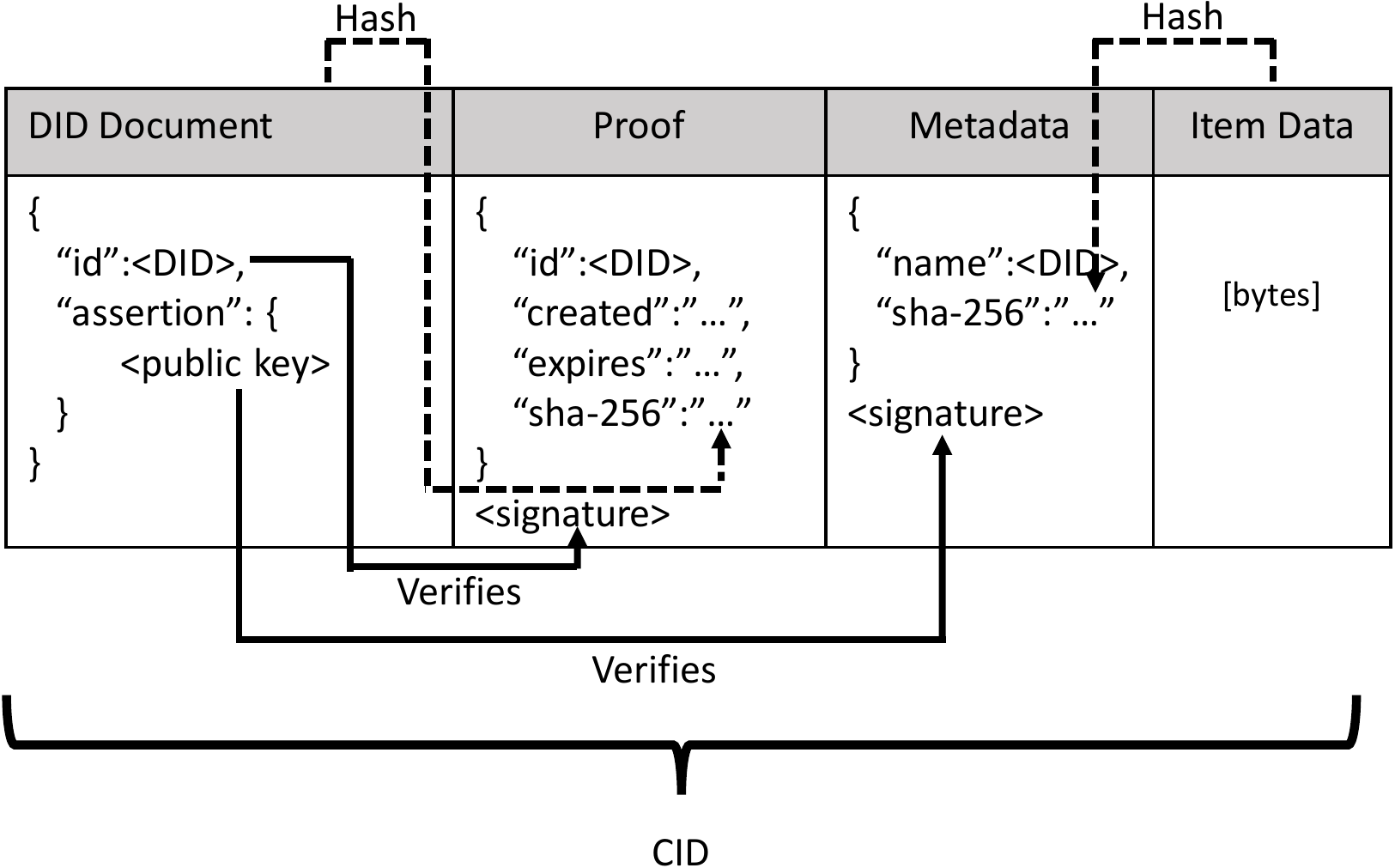}
  \caption{A self-verifiable content item.}
  \label{fig:svc}
\end{figure}

Initially, a content owner generates a self:did DID, and the corresponding DID document 
and proof. The DID is used as the content name. The generated 
DID document includes a public key (i.e., the \emph{assertion} property), which is used for signing some
\emph{metadata}. These metadata are a JSON string  that includes two properties: the content
\emph{name} and the content \emph{sha-256} digest. The final self-verifiable content item is a bundle
that includes the DID document, the proof, the signed metadata, and the item itself. The latter bundle is 
published in IPFS and receives an IPFS content identifier (CID): if any of the elements of this bundle
is modified the corresponding IPFS CID is also modified accordingly. Figure~\ref{fig:svc} illustrates an 
example of a self-verifiable content item. As it can be seen the DID is used for verifying the signature
of the proof, and the assertion key is used for verifying the signature of the metadata. Furthermore, 
the hash of the DID document is included in the proof and the hash of the item data is included in the
metadata.

An assertion key can be re-used for signing the metadata of multiple items. A content owner can rotate assertion keys
simply by changing the DID document and modifying the proof accordingly (since by changing the assertion key
the hash of the DID document also changes). Of course the updated self-verifiable item must be
``re-added'' to the IPFS network and receive a new CID.  

In order to map the generated DID (which is used as the content name) to the corresponding 
IPFS CID of the self-verifiable item we are using DNSlink. In particular 
the content owner creates in her DNS server a TXT record that maps the public key part of the did:self
DID to the appropriate content address. For simplicity reason we are assuming that this address is an IPFS CID,
but all other address types can be used as well. Every time the CID of the item is modified (e.g., the item data changes,
or the \emph{assertion} key is rotated) the corresponding
DNS record is updated. Therefore in order for a third party consumer to locate a content
item stored in IPFS using our naming system, he must know the DID of that item and the
domain name of the content owner. Given this information  the consumer can retrieve the 
corresponding content item and verify its authenticity using the process described in the following.

Initially, the consumer performs a domain name resolution and retrieves the appropriate DNS record: that
record is eventually translated into an IPFS CID. Then, the consumer retrieves the item that corresponds
to the CID from the IPFS network. The consumer extracts the DID document and the proof, and verifies
the validity of the DID document by executing the procedure described in the previous section. If the DID document is
valid the content authenticity can be verified by executing the following steps. 
\begin{enumerate}
    \item Verify that the \emph{name} property of the metadata includes the did:self DID.
    \item Verify that the  \emph{sha-256} property of the metadata includes the digest of the item's data.
    \item Verify the metadata signature using the key included in the \emph{assertion} property of the DID document.
\end{enumerate}

\begin{figure*}
  \centering
  \includegraphics[width=0.9\linewidth]{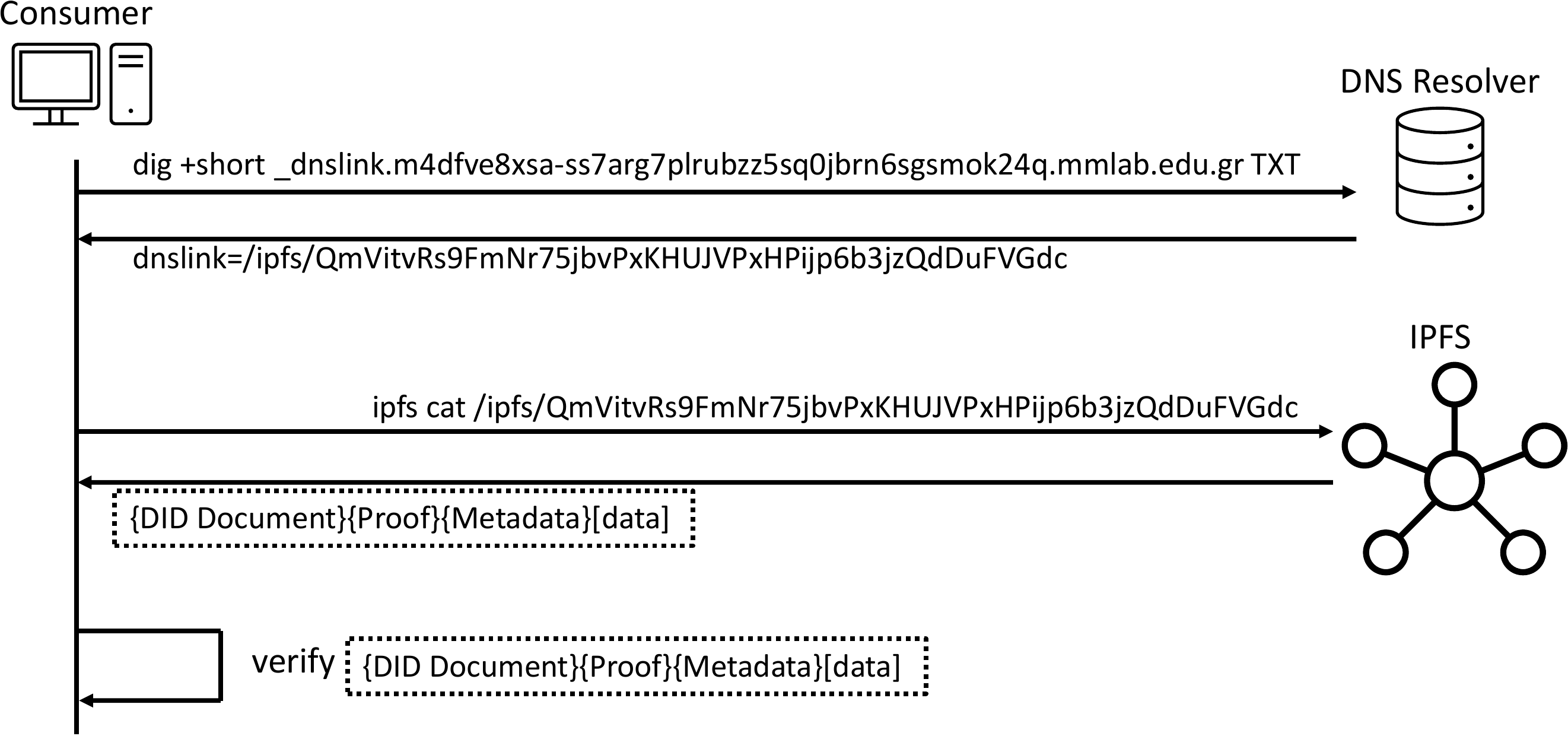}
  \caption{Retrieving a self-verifiable content item with DID did:self:m4dfve8xsa-ss7arg7plrubzz5sq0jbrn6sgsmok24q stored
  under the domain mmlab.edu.gr}
  \label{fig:ove}
\end{figure*}

An overview of this process is illustrated in Figure~\ref{fig:ove}. In particular, this figure illustrates the steps
required to retrieve and verify an item from the IPFS network, using a did:self-based name stored under the ``mmlab.edu.gr''
domain. It should be noted that except from the verification step, all other steps are transparently executed by the IPFS client application.
As it can be seen, initially a DNS resolution is performed and the DNSlink record is retrieved. The self-verifiable item's CID is 
extracted from the record and the item is retrieved from the IPFS network. Finally, the authenticity of the item is verified
using the information included in the item itself.

\subsection{Content storage delegation}
We are now revisiting the ``hosting service'' scenario presented in Section 1. Since our 
solution builds on DNSlink, content storage delegation can also be achieved by replacing
the TXT record of a DNSlink domain with a NS record that points to a DNS server controlled
by the hosting service. Nevertheless, and since we are focusing on mutable items, the hosting
service needs to be able to sign the metadata of the self-verifiable item
whenever the item is modified. Thanks to the use of DIDs this can be
easily and securely achieved as follows. The hosting service generates (or re-uses) a 
key pair which can be used for singing-verifying metadata signatures. The content owner
generates a new DID document that includes in the \emph{assertion} property the public
key of the hosting service. Then, the content owner sends the DID document and the proof
to the hosting service. The hosting service has not to (and cannot) modify the DID document
or the proof; it has only to include then in the self-verifiable content items. Nevertheless,
since the public key of the hosting service is included in the DID document, the hosting
service can generate valid metadata signatures.

When the content owner decides to change hosting service, she has simply to modify the
DNSlink record in her DNS server: the old hosting service cannot anymore ``disseminate''
the CID of the self-verifiable items. Furthermore, a content owner may set the expiration
property in the proof of the DID document sent to the hosting service. Therefore, the (old) hosting 
service not only it will not be able to disseminate CIDs but also, from a point on
it will not even be able to generate valid self-verifiable content items. 

\section{Implementation and Evaluation}
We have used the Python3 implementation of did:self.\footnote{https://github.com/mmlab-aueb/did-self-py}
Metadata are signed using JSON web signatures and they are serialized accordingly.
This functionality is implemented using the JWCrypto library.\footnote{https://jwcrypto.readthedocs.io/en/latest/}
SHA-256 hashes are calculated using Python's hashlib library. Our self-verified item
``generator'' script can either generate by itself a did:self DID, and the corresponding
DID document, proof, and assertion key, or these can be provided as input by the
content owner. This script also receives as input an existing file. Then the script
calculates the metadata, generates a digital signature, and creates a new file which
includes a line with the serialized DID, DID document, proof, metadata, and metadata signature, followed
by the contents of the input file. This new file is the self-verified version of the input
file, and it is then added in IPFS.

During the lifecycle of a self-verifiable item the following cryptographic operations
have to be performed. For the creation of the DID a user has
to generate an Ed22519 key pair, a DID document, and the corresponding proof. Additionally, she has to sign
the metadata property. For the item verification a user has to verify the DID document using the
provided proof, as well as the
signature of the metadata property.

Table 1 shows the time required (in ms) to perform the cryptographic operations of our system,
as measured in a desktop PC running Ubutnu 18.04, on an Intel i5 CPU, 3.1Ghz with 2GB of RAM.
As it can be seen most operations are executed in less than 3 ms.

\begin{table}
  \centering
  \caption{Cryptographic operations required by our system and their overhead.}
  \begin{tabular}{|c |c |} 

  \hline
  Operation & Time (ms) \\
  \hline
  Key pair generation & 46 \\
  DID document and proof generation & 2.7   \\
  JSON web signature calculation and serialization & 0.7 \\
  DID document verification & 1.5   \\
  JSON web signature verification & 0.2 \\
  \hline
 \end{tabular}
 \end{table}

 When it comes to storage overhead, Table 2 shows the size in bytes of the various components of a
 self-verifiable content item. It should be noted that the size of these components is constant and it
 is not affected by the size of the item data.

 \begin{table}
  \centering
  \caption{Size in bytes of the components of a self-verifiable content item.}
  \begin{tabular}{|c |c |} 

  \hline
  Component & Size (bytes) \\
  \hline
  DID document & 400 \\
  Proof & 420   \\
  Metadata + Signature & 134 + 110 \\
  \hline
 \end{tabular}
 \end{table}

\subsection{Security evaluation}
Our system is secure against attackers that have access to the metadata
signing key (i.e., the \emph{assertion} key), but cannot intervene to the DNS resolution
process. An attacker that has access to the assertion key of an owner can generate fake items. 
However since the attacker does not have access to the DNS server of the content owner
he cannot disseminate the fake items. In any case, a content owner can periodically 
rotate her assertion keys and use the \texttt{expires} property of the DID document proof.
Then, a breached assertion key can be used only for a limited time. 

An attacker that can intervene in the DNS resolution process, but does not have access to
the metadata signing key, can only redirect consumers to a fake
item or to an old version of the requested item. Both cases are DoS attacks, with the
latter having bigger impact since it cannot be easily detected by a consumer, whereas in the
former case the consumer will detect that the received item is fake. 

The ``old version of an item'' attack can be mitigated by creating a ``freshness'' indicator and 
having the consumers rejecting items or DNS replies that are ``old''. In the former case
(rejecting old items) a timestamp can be added to the metadata property of the self-verifiable
content item. In the second case, a timestamp can be added to the DNSlink TXT records 
and have that record signed with the assertion key. In both solutions the timestamp must
be updated periodically by the content owner no matter if the item has been modified. 
Although the first solution may appear simpler, it should be noted that every time the
timestamp included in the metadata property is updated, the self-certified item must be
``re-added'' to IPFS network and receive a new CID, then the corresponding DNSlink record
must be updated accordingly. 

\section{Discussion and conclusions}
In this paper we proposed a method for generating self-verifiable content items in IPFS.
Our solution supports mutable content items and leverages DNSlink to 
provide a mapping between the content name and its current IPFS content identifier (CID).

The main reason for selecting DNSlink to implement the latter mapping is compatibility with existing
IPFS tools. Indeed, using the available IPFS command line interface, any user can retrieve
a self-verifiable item from the IPFS network. Alternatively, our solution could have been
implemented using IPNS as follows: the did:self DID would be the IPNS name, and the IPNS
record would include, in addition to the content address, the DID document, the proof, and
the singed metadata; then the record would be signed using the ``assertion'' key of the
DID document. Using this approach, IPNS can be extended to support key rotation. 

Our solution achieves content authenticity protection without making any security
assumption about the security of DNSlink (and of DNS in general). In order to achieve
this property it sacrifices human readability. A trade-off between usability and security
could be the use of human readable names and the application of the ``trust on first
use'' principle. In particular our solution can be modify to use human readable domain
names that include a DNSlink record that ``points'' to the record of the actual did:self-based
DNSlink name. Assuming that the first time a client makes a DNS resolution for the human readable
name learns the correct DNSlink record, and that this record is cached permanently, human
readable domain names can be safely used. 

The presented solution does not take full advantage of the did:self method. The did:self 
method supports the notion of ``controllers'' which are in essence entities authorized to make
modifications the DID document. Using controllers it is possible to rotate the keys used
for signing a DID document proof, to use the same key for signing multiple DID document
proofs, as well as to securely delegate DID management to a $3^{rd}$ party. Similarly,
in the presented solution the assertion key is a public key but it can be as well a self:did
DID. In that case, self-verifiable content items must additionally include the DID document and the proof
that corresponds to the assertion DID. What we gain is that in scenarios such as our hosting service example,
it will be possible for the hosting service to rotate its metadata singing keys without
needing a new DID document from the content owner. 

Finally, although the work on this paper is focused on IPFS, self-verifiable content items
can be used in other systems as well. For instance, the project ``Self-Certified Names for 
Named-Data Networking''~\cite{scn4ndn} experiments with
a similar approach in the context of Information-Centric Networking paradigm. But even
legacy systems could benefit from our solution. For example, we believe that our approach
can be a more secure alternative to the ``subresource integrity'' 
HTML tag that is used for protecting the integrity of web resources loaded from a CDN; this tag
suffers from the same limitation as the CID IPFS identifier: every time the stored resource
is modified all web pages must change the integrity tag to the correct value. By storing
the resources in a self-verifiable format the integrity tag is not required any more
and the stored resources can be securely modified.

\section*{Acknowledgment}
This work was supported by a contract with the Waterford Institute of Technology under 
Article 15 of Grant Agreement number 871582 for financial support to third parties 
of EU H2020 project NGIatlantic.eu, a Research \& Innovation Action in the field of 
Next Generation Internet, and by a grant from Protocol Labs Inc. on ``Multi-Level DHT 
Design and Evaluation for IPFS.''

\bibliographystyle{IEEEtran}
\bibliography{IEEEabrv,references}
\end{document}